\documentclass[preprint,aps]{revtex4}
\usepackage{color}
\usepackage{mathtools}
\usepackage{diagbox}
\usepackage{appendix}
\usepackage{amsmath,amssymb,amsfonts,dcolumn,color,graphicx,graphics,latexsym,epsfig}
\usepackage{rsfso}
\usepackage[compatibility=false]{caption}
\usepackage{subcaption}
\usepackage{hyperref}
\usepackage{natbib}
\usepackage{float}
\usepackage{booktabs}

\begin{document}

\title{Novel triple barrier potential 
for axial gravitational perturbations of 
a family of Lorentzian
wormholes}

\author{Poulami Dutta Roy}
\email{poulamiphysics@iitkgp.ac.in}
\affiliation{Department of Physics, Indian Institute of Technology 
Kharagpur, 721 302, India}

\begin{abstract}
\noindent  
We study the behavior of a specific Lorentzian wormhole family under gravitational perturbations. In earlier work [EPJC 80, 850 (2020)], we have proved the stability of a test scalar field in the background of the wormhole family, where the effective potential was that of a double barrier. Continuing with the stability analysis, here we focus on the more physically relevant scenario, that of axial gravitational perturbations. Interestingly, we find that the effective potential is a triple barrier for lower angular momentum modes. This raises important questions on the ringdown of the corresponding wormhole geometry as well as the gravitational wave echo profile that we try to answer through our work. We study in detail how the geometry of each member wormhole affects the quasinormal modes, the time evolution of the signal as well as echoes which are, in general, very feeble in comparison to the main signal. Different `cleaning' techniques have been used to obtain the echo profile in the time evolution of the signal. Lastly, we dwell on the possibility of  
our wormhole family as a candidate black hole mimicker, as long as its stability is proven under all kinds of perturbations. We briefly present a comparison of the ringdown characteristics of these wormholes with that of a black hole, in support of this
speculation.
\end{abstract}


\maketitle

\newpage

\section {\bf Introduction}
\noindent Wormhole spacetimes have gripped the imagination of both the scientific community and the general public for decades ever since the term was coined by Wheeler \cite{wheeler_1957,wheeler_1955} to describe geometries that act as bridges between two universes. They were first discussed as early as 1916 in the pioneering work of Flamm \cite{flamm_1916} and then extended by Einstein and Rosen \cite{einstein_1935}. Their use as passageways for interstellar travel made wormholes a staple for science fiction. The physics community simultaneously took up the challenge of constructing an actual wormhole geometry and probing the possibility of their existence in nature. It turned out, in the seminal work of Misner and Thorne \cite{thorne_1988,morris_1988}, that such spacetimes need exotic matter to remain stable and traversable within the confines of General Relativity \cite{visser_2003,visser_1995,hochberg_1998} . If one is willing to explore alternate theories of gravity, then we get many scenarios where indeed wormholes can exist without the violation of energy conditions \cite{JLBS_2021,shaikh_2018,moraes_2019,reza_2017,bhawal_1992,maeda_2008,lobo_2009,dehghani_2009,kanti_2011,garcia_2010,bohmer_2012,harko_2013,kar_2015,bronnikov_2015,shaikh_2016,ovgun_2019,vagnozzi_2015}.  \\
\noindent Another hindrance one encounters while investigating wormholes is the issue of radial instability observed in some models. There have been number of stability studies on wormholes supported by scalar fields with the simplest being the Ellis-Bronnikov geometry.  Unfortunately, these models have been found to possess a growing mode that perturbs the throat to rapidly collapse or expand within a finite time thus stopping any form of passage through it \cite{shinkai_2002,gonzalez_2008,gonzalez_2008_1,bronnikov_2012}. Specifically, in \cite{shinkai_2002}, the dynamical instability of the Ellis-Bronnikov wormhole, which arises as a special case ($n=2$) in our wormhole family, has been studied. However, the question of stability of the other wormholes in the family considered here (i.e. those with $n\neq 2$) is yet to be addressed. These results have motivated research in finding the possibility of stable wormhole configurations through different avenues. Some examples include using alternate theories of gravity \cite{kanti_2011,cuyubamba_2018,kokubu_2015}, introducing certain number of Klein-Gordon scalar fields \cite{sarbach_2012}, considering source fluid with specific equation of state \cite{bronnikov_2013, novikov_2012}, using specific kind of exotic matter with thin shell \cite{garcia_2011} -- all 
providing interesting but very specific examples of wormhole geometries. Thus a robust wormhole model stable under all perturbation scenarios seems yet to be discovered. This highlights the significance and need for exploring the stability of various wormhole models under different perturbation schemes. \\
\noindent In spite of these issues plaguing wormholes, they are taken as serious contenders in black hole mimicker analysis.
Recently, after the detection of gravitational waves \cite{ligo_2016,ligo_2021}, wormholes are being considered as a possible black hole mimicker candidate along with a host of other exotic compact objects (ECO) \cite{cardoso_2016,nathan_2020,lemos_2008,cardoso_2019}. The huge array of ECOs which are horizonless include gravastars \cite{mazur_2004,visser_2003_gravastar}, boson stars \cite{macedo_2013,macedo_2016,olivares_2020}, quasiblack holes \cite{lemos_2004,lemos_2008,lemos_2007}, fuzzballs \cite{mathur_2005,ikeda_2021} and also wormholes \cite{mazza_2021,cardoso_2016,konoplya_2016}. For wormholes, there has been numerous studies where different aspects of such spacetimes have been looked at for their ability to mimic the behavior of a black hole \cite{shaikh_2018_lensing,banerjee_2019,cardoso_2016,tsukamoto_2012,ohgami_2015}. 
We will be particularly interested in the ringdown behavior of the wormholes involving their quasi-normal mode spectrum \cite{cardoso_2016,salcedo_2018,konoplya_2016,konoplya_2018,konoplya_2005}. The characteristic complex frequencies, dominating the ringdown stage, depend only on the source's parameters which makes them an excellent probe to deduce the stability and nature of the associated object. 
The ringdown spectrum can also be used as a tool to distinguish the wormholes from black holes which we will discuss in our work as well.
Apart from stability analysis, the QNMs associated with a wormhole geometry has been used to determine its shape \cite{konoplya_2018} and have been connected to the shadow radius produced by wormhole \cite{jusufi_2021}. \\
\noindent A further distinguishing feature associated with ECOs are the presence of echoes in the ringdown signal \cite{bueno_2018,cardoso_2016_echoes, mark_2017,micchi_2021,maggio_2019,maggio_2020}. There have been multiple studies on the formation of echoes in various spacetimes especially for wormholes and wormhole-blackhole transitions \cite{bronnikov_2020,churilova_2019}. Even black holes can have echoes too due to presence of additional structure near the horizon, hence making echoes a probe for the strong gravity regime \cite{abedi_2017,oshita_2020,damico_2020,oshita_2018,cardoso_2019_1,buoninfante_2020}. Therefore, detecting echoes in gravitational wave observations will always be associated with new physics either at the near-horizon region or near a compact object's surface. In fact recently, many studies have focused on finding traces of these echoes from the LIGO as well as Virgo data \cite{conklin_2017,abedi_2018,tsang_2019,uchikata_2019}. We will encounter these echoes in our work as well. Other detection channels for wormholes include gravitational lensing analysis \cite{abe_2010,godani_2021} and also looking for orbital perturbations caused by objects on other side of the wormhole throat, as formulated recently in \cite{dai_2019,simonetti_2020}.\\

\noindent With these motivations in mind, we study the behavior of a two parameter family of wormhole spacetimes under gravitational perturbations. Investigating the relationship between the QNMs and the wormhole's `shape' (geometry) is the prime objective of this work. A similar analysis of the wormhole family was done for scalar perturbations in our earlier work \cite{pdr_2020} which also included a detailed discussion of its matter content and energy conditions satisfied. In our present work, we perturb the metric itself, which can be split into axial and polar components \cite{chandrasekhar_1985,maggiore_2018} and is physically more relevant in the context of gravitational wave detection. We focus only on the axial part because of its simplicity  as well as the fact that matter remains unperturbed for our wormhole geometries. In the absence of an exact form of well-established matter sourcing the wormhole geometries, we use Einstein's equations to get the form of effective energy-momentum tensor which is interpreted as an anisotropic fluid. Following \cite{chen_2019}, we then find the matter sector remaining unperturbed indicated by a vanishing perturbation of the energy-momentum tensor under axial metric perturbations. 
Recently, there have been attempts to formulate matter sources that generate our wormhole family. For example,  wormhole solutions can be realized in the context of asymptotically safe gravity without any need for exotic matter, and these solutions are part of our wormhole family as shown in \cite{nilton_2022}. Our wormhole solutions can also be realized in $f(R)$ extended gravity where anisotropic dark matter sources them \cite{muniz_2022}. Both methods have found results that further motivate studies to investigate the nature of matter sourcing our wormhole family conclusively. For now continuing with the stability analysis, we calculate the corresponding quasi-normal modes associated with each member of the wormhole family. Such analysis will contribute towards developing proper templates for possible detection of wormholes through future gravitational wave observation.\\
A novel feature that emerged during our studies is the triple barrier effective potential associated with the perturbation equation for our wormholes. To this author's knowledge, such a scenario has not been observed in this context. This is true only for the lower angular momentum modes. For higher modes though, the potential becomes a single barrier. It is intriguing to study the echo profiles generated by the triple barrier because they are feeble and not observed directly in the time evolution spectrum. Special `cleaning' methods \cite{ghersi_2019,pdr_2020} have to be employed in order to visualise the echoes generated.\\
The paper is organised as follows. In Section II, we review the essential characteristics of our wormhole family. An in-depth analysis of these features has been done in our previous work. We then move on with the stability analysis under axial perturbations in Section III. The dependence of the triple barrier effective potential on different parameter values of the metric is discussed as well. Then, in Section IV, we calculate the quasi-normal modes and gravitational wave echoes as produced by the effective potential barrier. Finally, in Section V we study the possibility of this wormhole family being a viable black hole mimicker candidate by comparing the ringdown profiles with that of a black hole. We end with a brief discussion of our observations in Section VI.

\section {\bf Review of the wormhole family and its properties} 
\begin{figure}[ht]
      \centering
      \includegraphics[scale=1]{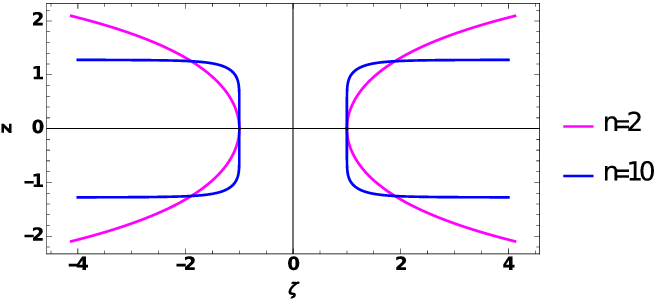}
      \centering
      \caption{Embedding diagram of a $t=constant$, $\theta=\pi/2$ spatial slice in 3-D Euclidean space for different values of $n$ and $b_0=1$. The cylindrical coordinates $z(\ell)$ and $\zeta(\ell)$ are plotted with $\ell$ being the parameter.}
      \label{fig:embed}
  \end{figure}

\noindent We will begin with a brief overview of the wormhole spacetime we are interested in. The well-known reflection-symmetric Ellis-Bronnikov wormhole, that was proposed independently by Ellis \cite{ellis_1973} and Bronnikov \cite{bronnikov_1973} in their 1973 papers, is of the form
\begin{align}
    ds^{2} = -dt^{2} + \frac{dr^{2}}{1-\frac{b_{0}^{2}}{r^{2}}}+ r^{2} d\theta^{2} + r^{2} \sin^{2}\theta d\phi^{2}
    \label{eq:ellis_metric}
\end{align}
where $b_0$ denotes the throat radius of the `drainhole'. Such a spacetime is geodesically complete, spherically symmetric, horizonless and is formed by a massless scalar field with a negative kinetic energy term, thereby making the matter threading the wormhole `exotic'. This is one of the simplest example of an  ultra-static form of Morris-Thorne wormhole \cite{thorne_1988}. A similar wormhole spacetime but with a more general form of the shape function was suggested in \cite{kar_1995} where the authors studied the transmission resonances associated with a line element 
\begin{align}
     ds^{2} = -dt^{2} + \frac{dr^{2}}{1-\frac{b(r)}{r}}+ r^{2} d\theta^{2} + r^{2} \sin^{2}\theta d\phi^{2}
     \label{eq:gen_ellis_r}
\end{align}
where $b(r)$, the shape function, is given by
\begin{gather}
   b(r)=r-r^{(3-2n)} (r^n-b_0^n)^{(2-\frac{2}{n})}.
    \label{eq:shape_function}
\end{gather}
The form of the shape function can be attributed to the relation assumed between the tortoise coordinate $\ell$ and the radial coordinate $r$,
\begin{align}
    r(\ell) = (\ell^{n}+b_{0}^{n})^{1/n}.
    \label{eq:radial_coord}
\end{align}
Using $\ell$ one can rewrite the metric in eq.(\ref{eq:gen_ellis_r}) in a simpler form as
\begin{align}
    ds^{2} = -dt^{2} + d\ell^{2}+ r^{2}(\ell) d\theta^{2} + r^{2}(\ell) \sin^{2}\theta d\phi^{2}.
    \label{eq:gen_ellis_metric}
\end{align}
Thus, we end up with a geometry that denotes a two parameter family of wormholes where the parameter $n$ is assumed to take only even values to ensure the smooth behavior of $r(\ell)$ over the entire region of $\ell$ ($-\infty < \ell <\infty$). The geometry (shape) of the wormhole is controlled by $n$ while the other parameter is simply the throat radius $b_0$. Substituting $n=2$ in the above metric gives back the Ellis-Bronnikov geometry while $n>2$ values correspond to distinct new wormholes that can be visualized through their respective embedding diagrams (fig.(\ref{fig:embed})). In our earlier work, which from now on, we will refer to as Paper-I \cite{pdr_2020}, we had given an in-depth analysis of this wormhole family including its geometry, matter, embedding diagrams, the behavior of $r(\ell)$ and the scalar quasi-normal modes associated with each member wormhole. For completeness we will review here, some of the salient features of the $n>2$ wormhole geometries in this work as well.\\

\noindent While we know the $n=2$ wormhole i.e. the Ellis-Bronnikov case is supported by a massless phantom scalar field, we are yet to understand fully the nature of matter threading the $n>2$ geometries. It is known that in GR we need exotic matter, atleast at the throat, to support a traversable wormhole \cite{visser_1995}. This necessary violation of energy conditions become a major drawback for the wormholes to exist within the confines of GR. Even so, Visser argues that violation of WEC does not necessarily imply the non-existence of wormholes in nature but simply a possibility of new physics that needs to be explored \cite{bekenstein_2013,barcelo_2002,santiago_2021}. In our wormhole family, all $n>2$ geometries show violation of WEC not at the throat but a small distance away from it. \\
\noindent This behaviour can be understood by writing the energy-momentum tensor components as a sum of contribution from the phantom scalar field and an additional matter term as shown in eq.(10), (11) and (12) of Paper-I. The nature of this extra matter term is not yet known but it was found to satisfy Averaged Null Energy Condition (ANEC). As a consequence, the flaring out of the geometry for $n>2$ wormholes happens a bit away from the throat. It can be contrasted with that of the $n=2$ wormhole where the flaring happens just from the throat as shown in fig.(\ref{fig:embed}). The $n>2$ geometries can thus be called as \textit{`long-necked' wormholes.} \\
 \noindent We also discussed the quasi-normal modes and echoes associated with each member of the family under the propagation of a massless scalar field. The stability of all wormholes of this family were proven under scalar wave propagation as observed in the damped signal of time domain evolution. After confirming its stability for the scalar case, we are now in a position to continue the stability analysis of the wormhole family under the more physically relevant gravitational perturbations.

\section{\bf Linear, axial gravitational perturbations}
\noindent In this section we will study the behavior of each wormhole geometry and their stability under gravitational perturbations. We follow the method used by Chandrasekhar \cite{chandrasekhar_1985} for deriving the perturbation equations but the derivation can as well be performed using the Newman-Penrose formalism \cite{newman_1961}. The formalism is mainly useful for a Petrov type-D spacetime which is a solution of the vacuum Einstein equation like the Kerr black hole. In this method the Weyl scalars are computed followed by the perturbation equations which can be separated into radial and angular parts. The situation gets complicated in presence of matter which is generally the case for wormhole spacetimes. Recently in \cite{matos_2021}, the authors have calculated the radial master equation for a general wormhole spacetime using the Newman-Penrose method and the Regge-Wheeler gauge. This analysis is applicable for spacetimes where the matter remains unaffected under axial perturbations thus excluding spacetimes that are solutions to Einstein-Maxwell equations. We will see that the final form of the radial equation (eq.(32)) and effective potential (eq.(33)) in \cite{matos_2021} matches exactly with our calculation, hence verifying the validity of our analysis.\\

\noindent Going back to our channel of analysis, we have the background metric of eq.(\ref{eq:gen_ellis_r}) which is spherically symmetric and static but in the perturbed state it might not remain so. Following Chandrasekhar \cite{chandrasekhar_1985}, we use a non-stationary but axisymmetric metric to denote the perturbed state of the wormholes. The line element denoting such a geometry is of the form,
\begin{align}
    ds^2 = - e^{2 \nu} \, dt^2 + e^{2\psi} (d\phi - q_r dr -q_\theta d\theta - \sigma dt)^2 + e^{2 \mu_r}dr^2 + e^{2 \mu_\theta} d\theta^2.
    \label{eq:pert_metric}
\end{align}
The components of the metric are functions of ($t, r, \theta$) but not $\phi$, due to axial symmetry. Also, as is evident, since the metric components depend on $t$, the perturbed object will evolve with time as it should. In the unperturbed state, it reduces to the background wormhole geometry, with the metric functions given as,
\begin{gather}
    e^{2 \nu}=1,\,\, e^{2\psi} = r^2 \sin^2\theta,\,\, e^{2 \mu_r} = \frac{1}{1-b(r)/r},\, \, e^{2 \mu_\theta} = r^2 \\
    q_r=q_\theta=\sigma=0.
\end{gather}
We notice that a perturbed state of our wormhole geometry will correspond to non-zero values of the quantities $\sigma, q_r$ and $q_\theta$ while $\nu, \mu_r, \mu_\theta, \psi$ will have small increments $\delta\nu, \delta\mu_r, \delta\mu_\theta, \delta\psi$. The non-zero value of the first set of metric quantities lead to presence of cross-terms which are of odd parity. This corresponds to a dragging of the frame of the perturbed object thus denoting rotation and is called axial perturbation. We will see that for our wormhole family the matter sector remains unaffected (shown in appendix).
On the other hand, the small increments $\delta\nu, \delta\mu_r, \delta\mu_\theta, \delta\psi$ do not impart any rotational effects and correspond to even parity. The matter content of the spacetime is indeed affected by such polar perturbation modes and while dealing with them, perturbations of the matter fields need to be taken into consideration.\\
In this work, we will focus on the axial perturbation modes mainly because of their simplicity.
The perturbation equations for any general Morris-Thorne wormhole geometry has been derived in multiple works in literature \cite{kim_2008,bronnikov_2012,kim_2004}. We will not repeat the entire calculations here but mention the important equations and results. \\
For the axial case, we have 
\begin{align}
    \bar{R}_{\mu \nu}  = 0
    \label{eq:perturbation}
\end{align}
where the Ricci tensor $\bar{R}_{\mu \nu}$ corresponds to that of the perturbed metric shown in eq.(\ref{eq:pert_metric}). A detailed calculation of the derivation of eq.(\ref{eq:perturbation}) is shown in the appendix. The perturbation equation mentioned above holds true only for the components relevant for axial case as in general the spacetime has matter and the RHS will not be 0.\\
Considering the $r \phi$ and $\theta \phi$ components of eq.(\ref{eq:perturbation}) and simplifying them using the ansatz $Q(t,r, \theta) = Q_{r \theta}\,\, \sin^3\theta\,\, r \sqrt{r^2 -b(r) r}$ with $e^{-i \omega t}$ as time dependence we get
\begin{align}
 \frac{\sin^3\theta}{r^2}\, \frac{\partial}{\partial \theta} \Big( \frac{1}{\sin^3\theta}\, \frac{\partial Q}{\partial \theta} \Big)\, + r \sqrt{r^2 - b(r) r}\, \frac{\partial}{\partial r} \Big( \frac{\sqrt{r^2-b(r)r}}{r^3}\, \frac{\partial Q}{\partial r} \Big) + \omega^2 Q = 0
\end{align}
where $Q_{r \theta} = q_r,_\theta - q_\theta,_r$. Thus, the quantity $Q_{r \theta}$ involves the derivatives of the metric elements $q_r$ and $q_\theta$.\\
\noindent In order to be able to separate the above equation into its radial and angular parts, we define $Q(r,\theta) = R(r) C_{m+2}^{-3/2} (\theta)$ so that
\begin{align}
    \Big(-\frac{(m+2)(m-1)}{r^2} R + r \sqrt{r^2 - b(r) r} \frac{d}{dr}\Big( \frac{\sqrt{r^2-b(r)r}}{r^3} \frac{dR}{dr} \Big) + \omega^2 R \Big) C_{m+2}^{-3/2} (\theta) = 0.
\end{align}
The parameter $m$ denotes the angular momentum mode arising due to the separation of variables. We can make a comment regarding the allowed values of $m$ by remembering the relation between the Gegenbauer polynomial and the Legendre polynomial $P_m(\theta)$ 
\begin{align*}
 C_{m+2}^{-3/2} (\theta) =  (P_{m,\theta\theta}- P_{m,\theta} \cot\theta ) \sin^2\theta.
\end{align*} 
One notices that for $m=0,1$ the Gegenbauer polynomial is identically 0. So, only for $m \geq 2$ one gets non-zero values and as the above equation must hold for all values of angular momentum modes, the radial master equation for axial perturbation will be 
\begin{align}
    r \sqrt{r^2 - b(r) r} \frac{d}{dr}\Big( \frac{\sqrt{r^2-b(r)r}}{r^3} \frac{dR}{dr} \Big) + \Big(\omega^2-\frac{(m+2)(m-1)}{r^2} \Big) R = 0.
    \label{eq:radial_eq}
\end{align}
Thus we have finally arrived at the master equation for the radial component of axial perturbation for our wormhole family. We can further simplify this by implementing the relation $R(r) = r Z(r)$, and using the tortoise coordinate we get the familiar Schr\"{o}dinger-like form of the radial equation,
\begin{align}
    \frac{d^2 Z}{d\ell^2} + \Big( \omega^2 -\frac{m(m+1)}{r^2} + \frac{5 b(r) -b'(r)r}{2 r^3} \Big) Z =0
    \label{eq:radial_eq_ell}
\end{align}
from which we can read off the effective potential \cite{kim_2008}
\begin{align}
    V(r) = \frac{m(m+1)}{r^2} + \frac{b' r -5b}{2 r^3}.
    \label{eq:potential}
\end{align}

\subsection{Effective potential: single/triple barriers}

\noindent The effective potential for our wormhole family can be obtained by substituting $b(r)$ of eq.(\ref{eq:shape_function}) into eq.(\ref{eq:potential}) so that we get the potential as
\begin{align}
    V(r) = \frac{m(m+1)-2}{r^2} + \frac{2 (r^n-b_0^n)^{1-2/n}}{r^n} - \frac{(n+1)\, b_0^n\, (r^n-b_0^n)^{1-2/n}}{r^{2n}}
\end{align}
while in terms of the tortoise coordinates it becomes
\begin{align}
    V(\ell) = \frac{m(m+1)-2}{(\ell^n+b_0^n)^{2/n}} + \frac{2 \ell^{n-2}}{(\ell^n+b_0^n)} - \frac{(n+1)\, b_0^n\, \ell^{n-2}}{(\ell^n+b_0^n)^2}.
\end{align}
We can reformulate the potential also in terms of dimensionless tortoise coordinate $x = \ell /b_0$ so that we get $V(x)$ in units of $b_0^2$
\begin{align}
    V(x) = \frac{1}{b_0^2} \Big( \frac{m(m+1)-2}{(x^n+1)^{2/n}} + \frac{2 x^{n-2}}{(x^n+1)} - \frac{(n+1)\, x^{n-2}}{(x^n+1)^2} \Big).
\end{align}

\noindent Once the form of the effective potential has been obtained, it can be plotted for different geometries and angular momentum modes for observing its behavior. We can study the dependence of V(x) on $x$ or alternatively we can take the throat radius as $b_0=1$ and plot $V(\ell)$ as a function of $\ell$.

\begin{figure}[h]
    \centering
    \includegraphics[width=0.6\textwidth]{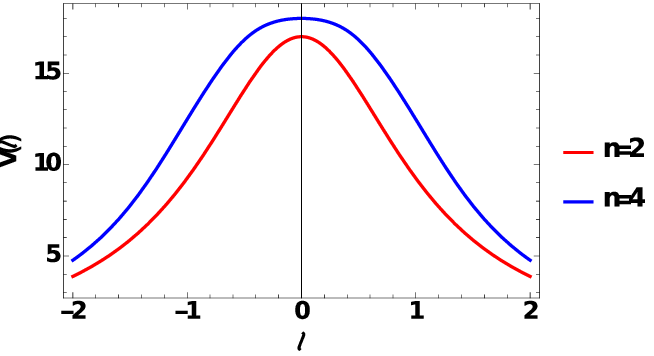}
    \caption{Potential showing single barrier for $n=2$ and $n=4$ geometries with $m=4,b_0=1 $.}
    \label{fig:pot_2_4}
\end{figure}

\begin{figure}[h]
 \centering
\begin{subfigure}[t]{0.38\textwidth}
  \centering
	\includegraphics[width=1.4\textwidth]{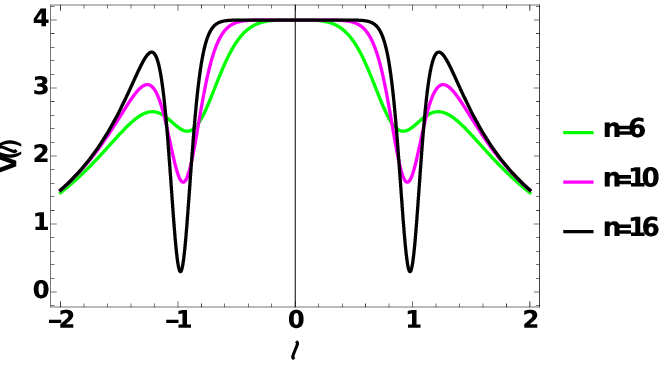}
 	\caption{ $m=2$}
 	\label{fig:pot_6_m2}
 \end{subfigure}
 \hspace{1in}
 \begin{subfigure}[t]{0.38\textwidth}
 	\centering
 	\includegraphics[width=1.4\textwidth]{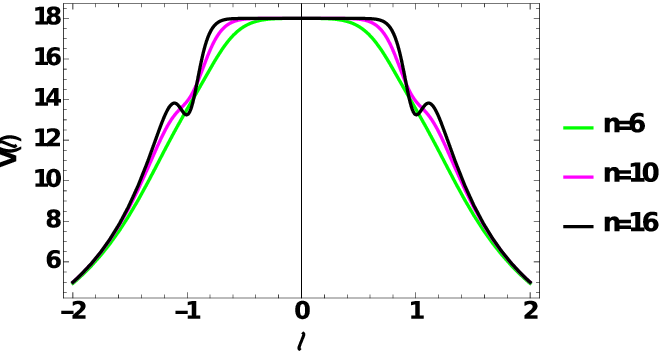}
 	\caption{$m=4$}
 	\label{fig:pot_6_m4}
 	\end{subfigure}
  \caption{Effective potential for $n=6,10,16$ where we find the triple barrier for lower $m$. For higher $m$ values the potential is strictly a single barrier.}
 \end{figure} 
 
\begin{figure}[h]
    \centering
    \includegraphics[width=0.6\textwidth]{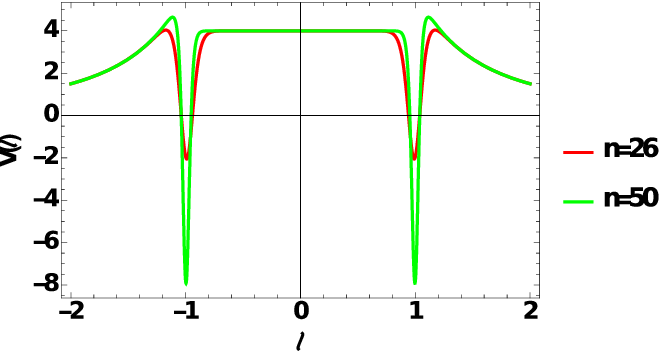}
    \caption{Triple barrier potential for higher $n$ geometries with $m=2$. The depth of the well increases with $n$ and becomes more negative.}
    \label{fig:pot_26}
\end{figure}

\noindent From the figures of the potential for different parameter values we can make certain observations which are summarised in the following three points.\\

\noindent $\bullet$ In fig.(\ref{fig:pot_2_4}) we find the potential to be a single barrier for wormholes corresponding to both $n=2$ and $n=4$ irrespective of the angular momentum mode. While the Ellis-Bronnikov wormhole always had a single barrier even for the scalar case (see Paper-I), the $n=4$ geometry, unlike its behavior under axial perturbation, had a double barrier potential just like its sister wormholes under scalar wave propagation.\\ 

\noindent $\bullet$ Moving on to $n>4$ geometries, we observe the effective potential to be characterized by triple barriers for lower angular momentum modes as can be seen in fig.(\ref{fig:pot_6_m2}). To the knowledge of this author, this is a novel behavior unique to our wormhole family. The height of the two symmetric peaks about the throat increases with increasing $n$. \\
The effective potential eventually becomes a single barrier for all geometries as we go to higher $m$ values. The two symmetric peaks about the central one merge to form a strict single barrier as shown in fig.(\ref{fig:pot_6_m4}). The exact value of `m' from which the potential becomes a single barrier goes on increasing as we go to higher `n' geometries. In contrast, for the scalar wave propagation case, it apparently seemed that the potential was a single barrier for higher $m$ values but in reality, on zooming in, the double barrier nature was found to be still prevalent as discussed in Paper-I.\\

\noindent $\bullet$ Finally, going to even higher values of $n$ (see fig.(\ref{fig:pot_26})), we find the triple barrier to be still present for lower $m$ modes but the depth of the well has now increased and is negative. It is known that negative potential wells may harbour bound states and hence have instabilities \cite{konoplya_2020}. So extra caution is required while studying the stability of the higher $n$ wormhole geometries.\\
Also, as mentioned above, with increasing $n$, the heights of the symmetric peaks about the throat increases. We find that for $n \geq 26$ wormholes, the two peaks become higher than the central one at the throat. Hence we choose $n=26$ for the plot shown in fig.(\ref{fig:pot_26}). For higher modes, the height of the two peaks gradually decreases, later merging into a single barrier.\\

\noindent In an attempt to derive more information about the extrema of the potential, especially for lower angular momentum modes, we again go back to dimensionless variable $x = \frac{\ell}{b_0}$ so that the points of maxima and minima for $V(\ell)$ will be a solution of the following equation,
\begin{align}
   x^{n-1} \Big[x^{-2}\Big(-4 x^{2n} +x^n\, (n^2+5n-6)+\,(3n-n^2-2)\Big) - \frac{2\, (m(m+1)-2)\, }{(1+x^n)^{\frac{2}{n}-2}} \Big] = 0.
\end{align}
We see that $x=\ell=0$ is always a point of extrema irrespective of geometry and $m$ value. For $m=2,3$, i.e. lower angular momentum modes, we can check whether the bracketed quantity gives any real roots for $n=4$ as they will correspond to points of extrema other than $\ell=0$. We find that for $n=4,m=2$ the quantity in bracket has only imaginary roots indicating that there are no points of extrema. Therefore, $n=4$ wormhole always has a single barrier similar to $n=2$ case. If one calculates for $n=6$, there will be two solutions other than $\ell=0$ indicating a maxima and a minima and hence a triple barrier for $m=2$. Unfortunately we cannot solve the equation for any general $m$ and $n$ value and hence can comment on each geometry ($n$ value) separately.\\


\section{\bf Stability analysis: Quasi-normal modes and gravitational wave echoes}

\noindent The ringdown of a perturbed object is a well studied phenomenon using linear perturbation theory. Due to the recent detections of gravitational waves there has been a huge interest in the study of the ringdown profile, which is dominated by the quasi-normal modes, for various objects like black holes, ECOs as well as neutron stars. Quasi-normal mode (QNM) frequencies are complex characteristic frequencies through which a perturbed object looses its energy and settles into an equilibrium state \cite{vishveshwara_1970_1,vishveshwara_1970_2,kokkotas_1999}. These modes are very significant because of their dependence only on the parameters of the final object and not on the cause producing them. As per our assumption, we have the time dependence as $e^{-i \omega t}$ with $\omega= \omega_r + i \omega_i$ where a negative $\omega_i$ will indicate damping and hence stability of the system over time. The QNMs apart from proving the stability of an object can also be used as a test for GR and other theories of gravity \cite{bhattacharyya_2017,konoplya_2016_1}.\\
 
\subsection{Numerical computation of QNMs}

\begin{figure}[h]
    \centering
    \includegraphics[width=0.5\textwidth]{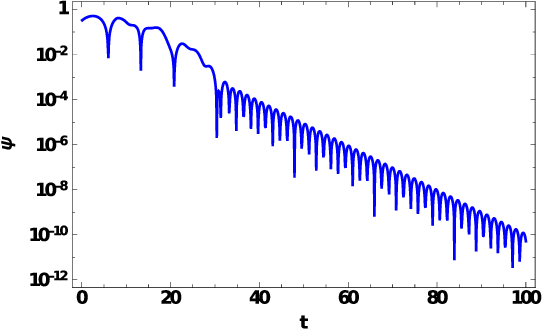}
    \caption{Time domain profile showing the characteristic QNM ringing in semi-logarithmic scale for $n=10,m=2$ as observed at $\ell=5$. The integration is done on a $u-v$ null grid with a step size 0.1. The initial conditions are defined on the lines $v=0$ as a Gaussian profile centered at $u=6$, $\psi(u,0)= e^{\frac{-(u-6)^{2}}{2 \times 4^2}}$ and, on the $u=0$ line, as a constant, $\psi(0,v)= constant$, the value of which is determined by $\psi(0,0)$. For details on the discretization scheme, see \cite{konoplya_2011}.}
    \label{fig:TD}
\end{figure}

\noindent We will calculate the QNMs for our wormhole family by solving the radial differential equation eq.(\ref{eq:radial_eq_ell}) numerically. The boundary conditions for a wormhole are similar to that of a black hole, but with outgoing waves at spatial infinity and at the throat. We use the direct integration method and the Prony extraction technique to obtain the dominant fundamental QNM frequencies from the time domain profiles. \\
In direct integration, the radial equation is integrated numerically by imposing proper boundary conditions at the throat and at spatial infinity. Since our wormhole has reflection symmetry about the throat we can separate the solutions into even and odd cases. The method is discussed in detail in \cite{pdr_2020,aneesh_2018}. \\
On the other hand, for implementing the Prony extraction technique we need to obtain the time evolution plots for different wormhole geometries. We begin by recasting the master radial wave equation (keeping the time dependence) as
\begin{align}
    \frac{\partial^{2}\psi}{\partial t^{2}} - \frac{\partial^{2}\psi}{\partial \ell^{2}} + V_{eff}(\ell) \psi =0 
    \label{eq:QNM}
\end{align} 
with $\psi$ denoting the perturbation. Writing the above equation in light cone coordinates ($du = dt- d\ell$ and $dv = dt + d\ell$) and
using a Gaussian pulse as initial condition along the $u$ and $v$ grid lines we numerically integrate
to obtain the time-domain profiles as shown in fig.(\ref{fig:TD}). The damped ringing in time, exhibiting the decay of the axial gravitational perturbation is clearly visible in
the plot. Once we obtain it, we can extract the most dominant frequency by fitting an exponentially damped signal to it. \cite{konoplya_2011} summarises the process in great detail along with the discretization scheme used for generating the plot. \\

\noindent Using the numerical methods discussed above we calculate the QNM frequencies associated with different wormhole geometries and for different angular momentum modes. All computations are done using \textit{Mathematica 12}. The tables (\ref{tab:n2}) and (\ref{tab:n4}) lists QNM frequencies calculated for $n=2,4$ and $n=6$. The values calculated from both methods show better matching for lower angular momentum modes. As $m$ value increases, DI becomes less stable for wide range of matching points. So, for further analysis we will prefer the results as obtained using the Prony method. 
\begin{table}[H]
\begin{minipage}[t]{0.5\linewidth}
\begin{tabular}{|c|c|c|}
     \hline
     m & Prony & DI \\ 
    \hline 
   2 &  1.73846 -i 0.305138 & 1.73769-i 0.305138\\ 
    
    3  & 2.95552 -i 0.409773 &  2.95241 -i 0.409983\\ 
    
    4 &  4.08393 -i 0.446586 & 4.07626 -i 0.44908\\
   
    5  & 5.6995 -i 0.463105 &  5.15483 -i 0.467286\\ 
    
    6 & 6.23495 -i 0.471021 &  6.20884 -i 0.477064\\
    \hline
\end{tabular}
\centering
\caption{\label{tab:n2} $\omega_{QNM}$ values for different modes for $n=2,b_0=1$. The values for $m=2,3,4$ can be verified from \protect\cite{salcedo_2018}.}
\end{minipage}
\hspace{0.5in}
\begin{minipage}[t]{0.5\linewidth}
\begin{tabular}{|c|c|c|}
    \hline
       m  & Prony & DI  \\
       \hline
       2 &  1.86183 -i 0.270285  & 1.86212 -i 0.270645\\
       3 & 3.14391 -i 0.316273 & 3.14398 -i 0.317279\\
       4 & 4.29188 -i 0.285576 & 4.28394 -i 0.287925\\
       5 & 5.37117 -i 0.255389 & 5.35662 -i 0.258602\\
       6 & 6.42508 -i 0.231619 & 6.4 -i 0.236127\\
       \hline
    \end{tabular}
    \caption{$\omega_{QNM}$ for $n=4$, $b_0=1$.}
    \label{tab:n4}
\end{minipage}
\end{table}
\begin{table}[H]
    \centering
    \begin{tabular}{|c|c|c|}
    \hline
       m  & Prony & DI  \\
       \hline
       2  & 1.89453 -i 0.247715 & 1.89443 -i 0.247615\\
       3 & 3.18616 -i 0.283299 & 3.18363 -i 0.283909\\
       4 & 4.33205 -i 0.234648 & 4.32457 -i 0.236658 \\
       5 & 5.40471 -i 0.193488 & 5.39035 -i 0.196454 \\
       6 & 6.45137 -i 0.164629 & 6.42671 -i 0.168574\\
       \hline
    \end{tabular}
    \caption{$\omega_{QNM}$ for $n=6$, $b_0=1$}
    \label{tab:n6}
\end{table}

\noindent We now study how the geometry of the wormholes influence the QNM values i.e. how the QNMs depend on parameter $n$.

\begin{figure}[h]
 \centering
\begin{subfigure}[t]{0.41\textwidth}
  \centering
	\includegraphics[width=1.2\textwidth]{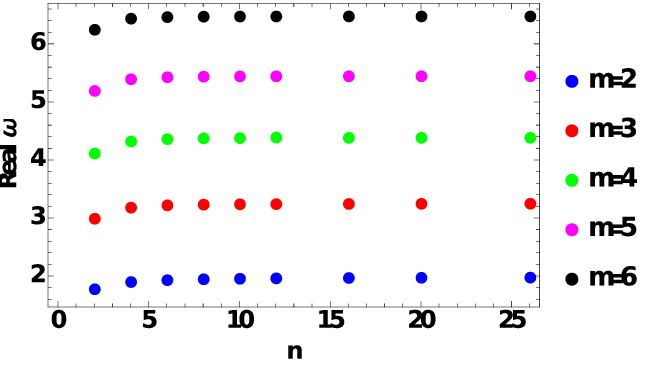}
 	\caption{ Variation of real part of QNM frequency with $n$.}
 \end{subfigure}
 \hspace{1in}
 \begin{subfigure}[t]{0.41\textwidth}
 	\centering
 	\includegraphics[width=1.2\textwidth]{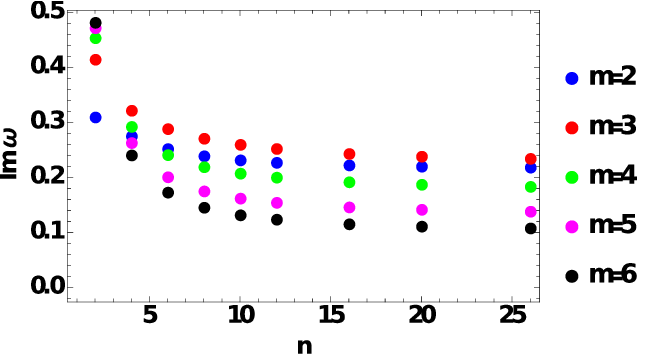}
 	\caption{Variation of imaginary part of QNM frequency with $n$.}
 	\end{subfigure}
 	\caption{QNMs are plotted with $m$ as parameter with $b_0=1$.}
 	\label{fig:QNM_plot}
 \end{figure}  
 
\noindent $\bullet$ The behavior of the real part of the QNM frequency for different modes and geometries are similar. As $m$ increases, the magnitude of the QNM increases indicating a higher value of frequency. This is because of the fact that with increasing $m$, height of the potential barrier also increases and so only waves with enough energy i.e. high frequency are able to cross the barrier.\\

\begin{figure}[h]
    \centering
    \includegraphics[width=0.5\textwidth]{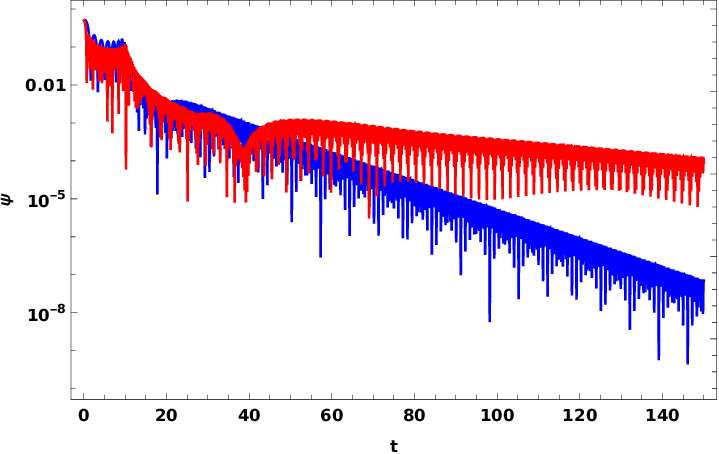}
    \caption{Time domain profiles for $n=10$, $m=8$ (blue) and $m=20$(red) as observed at $\ell=5$ and with the initial signal being a Gaussian wave $\psi(u,0)= e^{\frac{-(u-6)^{2}}{2 \times 4^2}}$. The long-lived QNMs are clearly visible as they dominate the ringdown for higher angular momentum modes resulting in slow decay of signal over time.}
    \label{fig:slow_decay}
\end{figure}

\noindent $\bullet$ The behavior of the imaginary part is rather different for $n=2$. As $m$ increases, the magnitude of $\omega_i$ increases, indicating that $m=2$ is the least damped mode. For $n>2$ geometries, the value of $\omega_i$ first increases and then decreases with $m$. So $\omega_i$ for $m=3$ generally has higher magnitude than all other $m$. Also we cannot predict the most dominant mode for $n>2$ geometries as the least damping time will correspond to $m \rightarrow \infty$. The fig.(\ref{fig:slow_decay}) shows the effect of the long-lived modes resulting in the slow damping of the signal over time. The red curve corresponds to $m=20$ which clearly decays very slowly than the $m=8$ mode for $n=10$ wormhole geometry.\\
A similar result is reported in \cite{liu_2021} for a completely different wormhole scenario. Hence, the above analysis highlights the fact that we can easily distinguish the wormholes of different `shapes' just from their fundamental axial quasinormal mode frequency which was one of the goals of this work.\\

\begin{figure}
    \centering
    \includegraphics[width=0.5\textwidth]{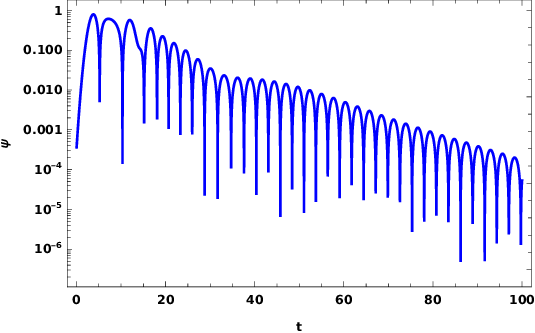}
    \caption{For $n=1200,m=2$ time domain plot showing evolution of an initial Gaussian signal with grid spacing $h=0.02$. The damped nature of the signal over time indicates stability of the geometry.}
    \label{fig:n1200}
\end{figure}

\noindent \textbf {Are higher $n$ wormhole geometries unstable?}\\
As we have seen in the potential plots, the minima for some cases reach negative values (large $n$, small $m$) (see fig.(\ref{fig:pot_26})), which might harbour bound states and may have instabilities. In all the previous time domain plots, we have used a grid spacing of $h=0.1$ with the convergence of the integration scheme being checked for smaller $h$ values. But for high $n$ geometries choosing $h=0.1$ shows instability in the time domain profile which is a numerical artifact as they disappear and we get damped signal for smaller $h$ values like 0.02 or 0.01. This happens because the well width is so narrow that the variation in the potential cannot be detected during the numerical analysis. Hence we need to reduce the grid spacing sufficiently so that it can trace the variations of the potential near the wells. No instability has been observed for any parameter values in the time evolution profiles of our wormhole family. The damped signal indicates the stability of our wormhole family under axial perturbation. For $n=1200$ we show the damped time domain signal in  fig.(\ref{fig:n1200}).\\

\subsection{Observing echoes in different wormhole geometries}

\begin{figure}[h]
    \centering
    \includegraphics[width=0.95\textwidth]{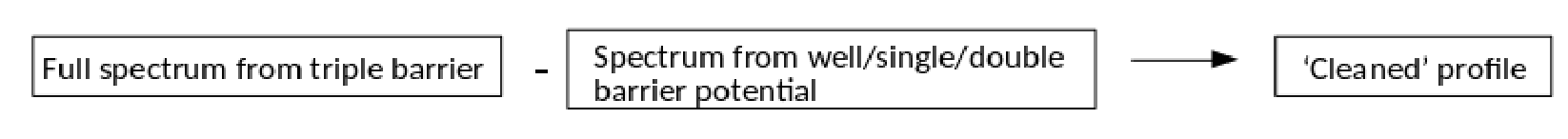}
    \caption{Block diagram showing the outline of the `cleaning' procedure.}
   \label{fig:block}
\end{figure}

\noindent In this section we will have a look at the echo profiles in the time domain signal generated by the triple potential barrier. Whenever a potential is characterized by a second peak or a reflecting surface, the gravitational wave signal not only decays over time but gets reflected. This leads to repetitive bursts of signal, after the initial ringdown, with decreasing amplitudes called `echoes'. Under scalar wave propagation our wormhole geometries had a double barrier and hence had distinct echo signatures in their time domain profiles. These echo profiles became more prominent for larger $n$ geometries because of the distinct reflection occurring from the sharp peaks in the potential. The separation between two echo signals is equal to the time taken by the signal to travel to-and-fro between the two peaks.\\
\noindent We expect to see similar echo signatures in the case of axial perturbation as well because here too we have triple potential barriers. It is needless to say that the generation of the echo signal for this case will be much more rich and complicated because the signal has to get reflected between multiple barriers. But the first question that we might ask is \textit{where are the echoes?} In fig.(\ref{fig:TD}) we see that even for $m=2$ case there are no echoes. Even if the time domain profile is calculated for longer time, echoes are not visible. One might think that the situation is similar to the scalar case (as discussed in Paper-I) where low $n$ wormholes did not show echoes due to the wide peaks causing back scattering and hence going to large $n$ solved the issue as the potentials had sharp and narrow peaks. But it is not so simple in the case of axial perturbations. Here too the back scattering from the wide tails of the triple barrier suppresses the echoes. To add to the problem, the separation between two consecutive peaks is very small and there is very less chance of observing an echo. The situation does not get better even if we go to higher $n$ geometries because for axial perturbation, with increasing $n$, the depth of the well (minima) increases but the height of the peaks remain almost unchanged and so does the width of the barriers. \\
One possible way out of this scenario is to `clean' the profile. Such a technique was used by us in Paper-I to observe echoes for small $n$ geometries following \cite{ghersi_2019}. The main essence of the process was to remove, from the full spectrum, the effect of the scattering of signal from the single barrier (see Fig.(\ref{fig:block})). After subtraction, one should ideally be left with only the scattering happening due to the double barrier i.e. echoes. But implementing such a method for the axial perturbation case becomes quite tricky. We first need to ascertain the effect of which potential peak needs to be subtracted from the full spectrum since now we have three distinct single barriers. We begin by taking a look at the three possible scenarios at hand for a specific wormhole geometry, say $n=10,m=2$, 
\begin{align}
    \begin{split}
        V(\ell) (single \,\,barrier) = \begin{cases} V_{eff} (\ell); & \, 0.955 \leq \ell \\
        1.6; & \,\, \ell < 0.955
    \end{cases}    
    \end{split}
\end{align}

\begin{align}
    \begin{split}
        V(\ell) (double\,\, barrier) = \begin{cases} V_{eff} (\ell); & \, -0.955 \leq \ell \\
        1.6; & \,\, \ell < -0.955
    \end{cases}    
    \end{split}
\end{align}

\begin{align}
    \begin{split}
        V(\ell) (well) = \begin{cases} V_{eff} (\ell); & \, 0 \leq \ell \\
        4; & \,\, \ell < 0
    \end{cases}    
    \end{split}
\end{align}

\noindent We take the single barrier to be the right most peak among the three barriers and so the potential is that of the wormhole from $\ell \geq 0.955$ which is the point of minima. For region $\ell < 0.955$, the potential is kept constant and equal to the value of the potential at $\ell=0.955$ i.e. 1.6. Similarly, the double barrier potential is considered by keeping the potential peaks at $\ell=0$ and the one to its right so that the potential is that of the wormhole for $\ell \geq -0.955$ while for $\ell< -0.955$ it is again constant with value 1.6. All these values correspond to the $n=10,m=2$ geometry. The well potential simply denotes the first well a signal encounters on the positive side of the throat so that for $\ell \geq 0$ we have the wormhole potential but on the other side the potential becomes constant and equal to the value at throat.
Before we move on with the subtraction procedure let us see how the spectrum for each of these potential cases behave.
\begin{figure}[h]
 \centering
\begin{subfigure}[t]{0.42\textwidth}
  \centering
	\includegraphics[width=1\textwidth]{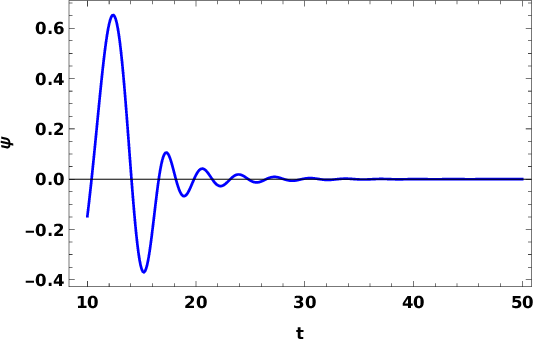}
 	\caption{ Full spectrum}
 \end{subfigure}
 \hspace{0.8in}
\begin{subfigure}[t]{0.42\textwidth}
  \centering
	\includegraphics[width=1\textwidth]{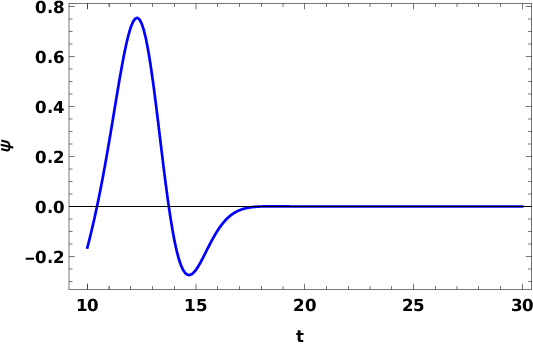}
 	\caption{ Spectrum from scattering off single barrier}
 \end{subfigure}
 \hspace{0.8in}
 \begin{subfigure}[t]{0.42\textwidth}
 	\centering
 	\includegraphics[width=1\textwidth]{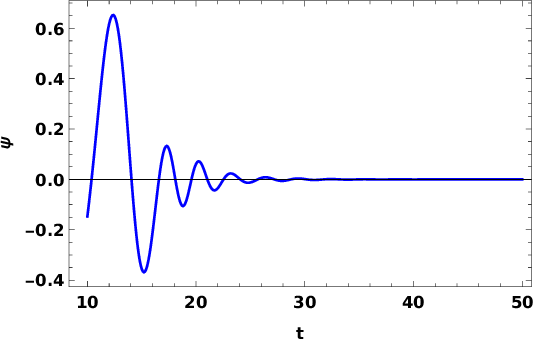}
 	\caption{Spectrum from scattering off double barrier}
 	\end{subfigure}
 \hspace{0.8in}	
 \begin{subfigure}[t]{0.42\textwidth}
  \centering
	\includegraphics[width=1\textwidth]{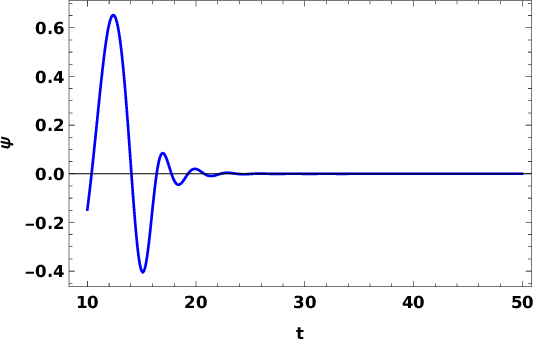}
 	\caption{Spectrum scattering off well}
 \end{subfigure}	
 	\caption{The time domain signals with initial Gaussian pulse $\psi = e^{-\frac{(\ell-4)^2}{2 \times 1^2}}$ observed at $\ell= 5$ for $n=10,m=2$ geometry.}
 	\label{fig:unclean}
 \end{figure} 

\begin{figure}[h]
 \centering
\begin{subfigure}[t]{0.4\textwidth}
  \centering
	\includegraphics[width=1\textwidth]{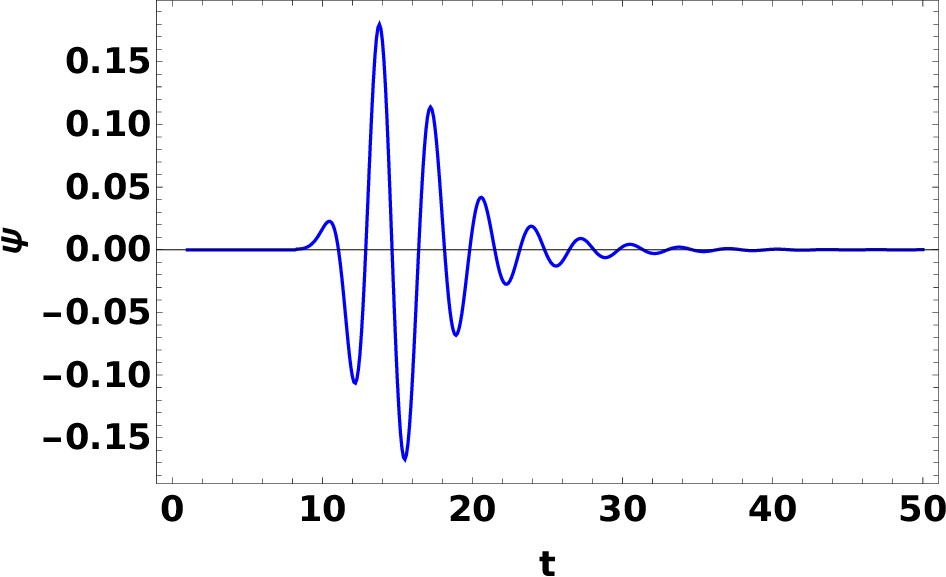}
 	\caption{Effect of single barrier subtracted from full spectrum}
 \end{subfigure}
 \hspace{1.5in}
\begin{subfigure}[t]{0.38\textwidth}
  \centering
	\includegraphics[width=1\textwidth]{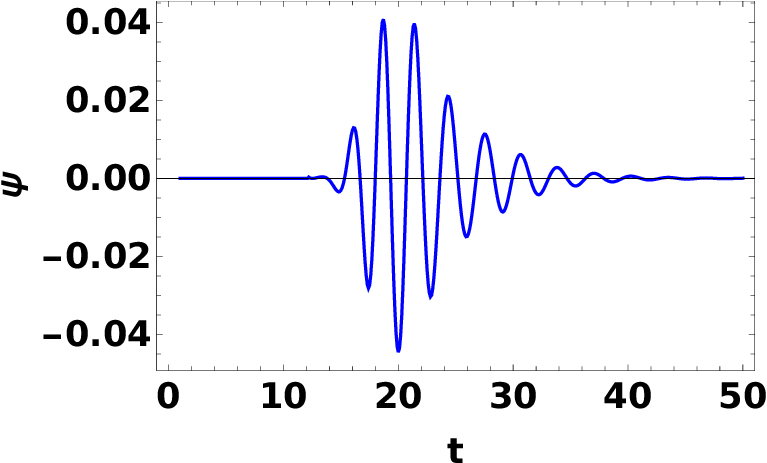}
 	\caption{ Effect of double barrier subtracted from full spectrum}
 \end{subfigure}
 \hspace{1in}
 \begin{subfigure}[t]{0.38\textwidth}
 	\centering
 	\includegraphics[width=1\textwidth]{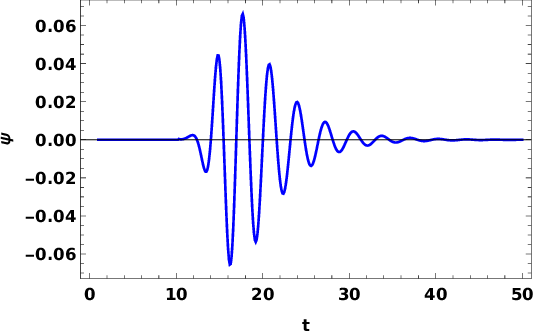}
 	\caption{Effect of the first well subtracted from full spectrum}
 	\end{subfigure}
 	\caption{Different `cleaning' techniques highlight the echo signal in the time domain profile which was otherwise hidden in the full spectrum.}
 	\label{fig:clean}
 \end{figure}

\noindent To make sure that after cleaning if we observe the echo, that is not the effect of solely the potentials that are subtracted, we plot the scattering from these potentials alone. It can be seen from Fig.(\ref{fig:unclean}) that none of the potential scenarios possess echoes but just a damped ringdown signal. We now subtract the effect of scattering from each of these potentials from the full spectrum. The `cleaned' profile corresponding to each of the above cases is shown in Fig.(\ref{fig:clean}). The echo signal can now clearly be seen after cleaning for all three cases.\\
\noindent In the third figure (Fig.(\ref{fig:clean}c)), we subtract the effect of the first potential minima or the well and observe an echo. The difference of this signal with subtracting the double barrier potential is that we can observe the scattering effect occurring from the tail of the central peak on the left. Finally, this procedure helps us clearly visualize the effect of the triple barrier on the echo signal. For such a potential the signal will not only be reflected between the first two peaks but some part will be transmitted into the other potential gap. There again it will undergo multiple reflection and some part will be transmitted to infinity while some will again get transferred to the other potential gap. In this way the echo signal for the triple barrier is a superposition of multiple echoes produced via reflections between the two potential gaps.\\
From the figures showing echo profiles, we see that the height of the signal increases with time and then decreases, indicating the profile to be an `echo'. Also, there is no initial ringdown because it has been subtracted and we are left only with the effect of reflections from the potential barriers.\\
We now continue with some interesting observations which can be made from the echo profiles.\\
 
\noindent $\bullet$ \textbf {Which peak, out of the three, should be chosen as the single barrier?}\\
The single barrier that we used for `cleaning' the profile and observe echoes is taken as the first peak of the potential on the right of $\ell=0$ that a signal from infinity will encounter first. We have also checked with the central peak but the echoes are not distinct after cleaning. This may be due to the fact that we are taking constant potentials on both sides of the peak which changes the behavior of the potential at both infinities.\\

\noindent $\bullet$ \textbf {Why is only one echo signal observed?}\\
It is very difficult to observe multiple echo packets because the signal damps very rapidly and the amplitude of the signal is too low to observe echoes even with cleaning procedure. Thus all the reflections between the different potential peaks superimpose to form the dominant echo signal with significant amplitude, which we observe.\\

\begin{figure}
    \centering
    \includegraphics[width=0.8\textwidth]{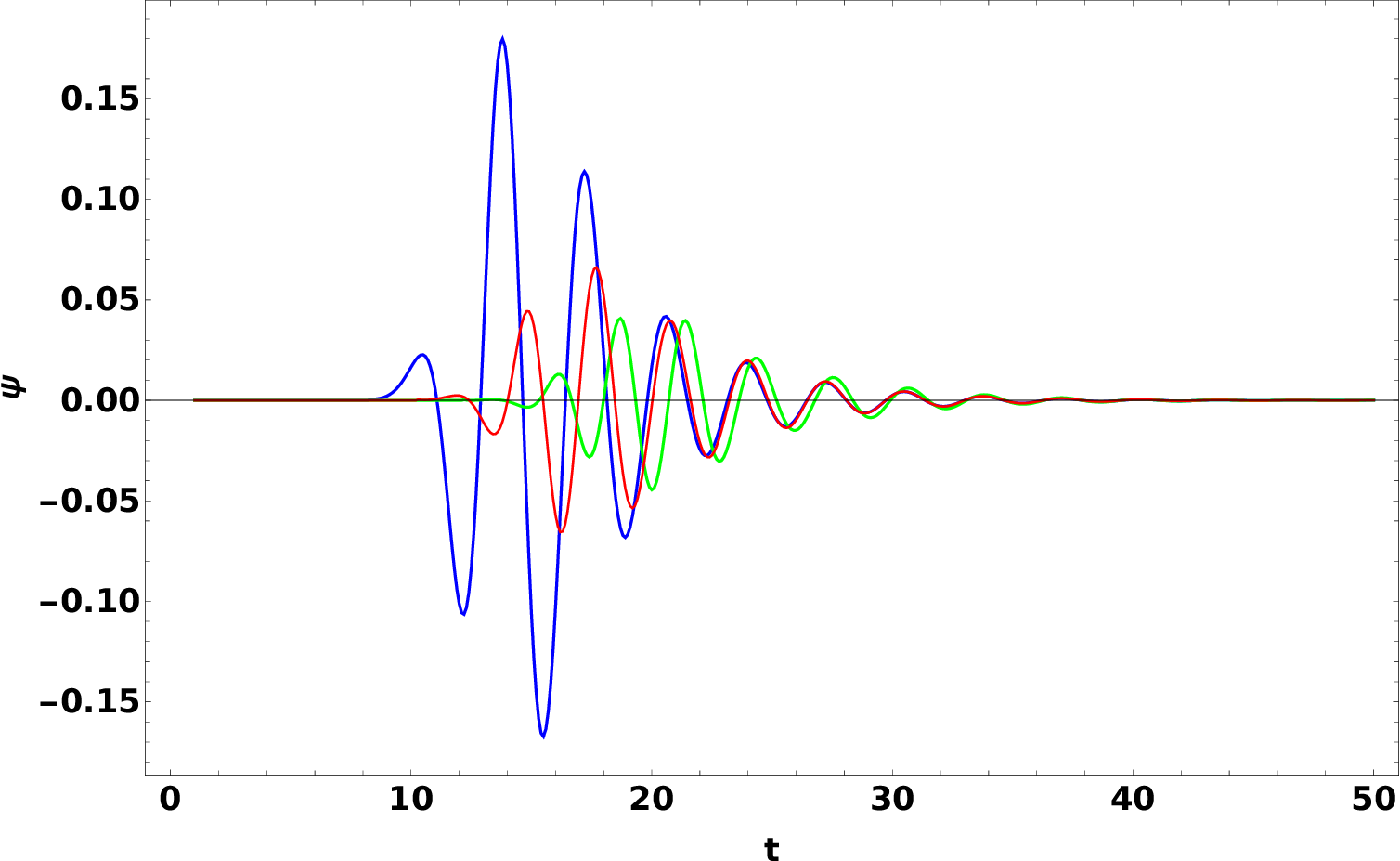}
    \caption{For $n=10,m=2$ echo profiles obtained by different potential subtraction; blue (full-single), green (full-double), red (full -well)}
    \label{fig:n10m2}
\end{figure}

\noindent $\bullet$ \textbf {Comparison of signal from different cleaning procedures}\\
In fig.(\ref{fig:n10m2}) we observe the echo profiles obtained by subtracting different sections of the potential for $n=10$ wormhole. The blue curve denotes the echo observed while subtracting just the effect of the first single peak. Hence the signal starts at an earlier time as well. Now, when we subtract the effect of the well potential we get the red curve which denotes that the echo starts a bit late and is of weaker strength. Finally, we get the green curve corresponding to the echo obtained by subtracting the effect of the double barrier that starts even later and is more weak than the other two.\\

\begin{figure}
    \centering
    \includegraphics[width=0.8\textwidth]{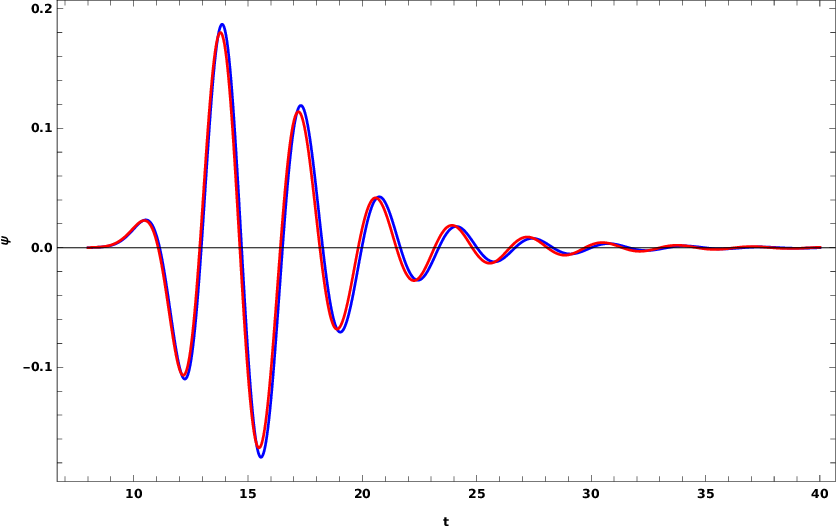}
    \caption{The echo profile obtained by removing the effect of the single barrier for $n=6$ (blue curve) and $n=10$ (red curve) geometries with $m=2$.}
    \label{fig:n6n10}
\end{figure}

\noindent $\bullet$ \textbf {Why echo amplitude decreases with increasing $n$?}\\
With increasing $n$ we find the echo amplitude to be decreasing as seen in fig.(\ref{fig:n6n10}). This occurs because of the increasing depth of the well for higher geometries which makes it more difficult for a trapped wave to escape from the well. So even though the peak height is small and the potential peaks have wide tails for small $n$, we still get higher echo amplitude after `cleaning' because of the small well depth. \\

\begin{figure}
    \centering
    \includegraphics[width=0.6\textwidth]{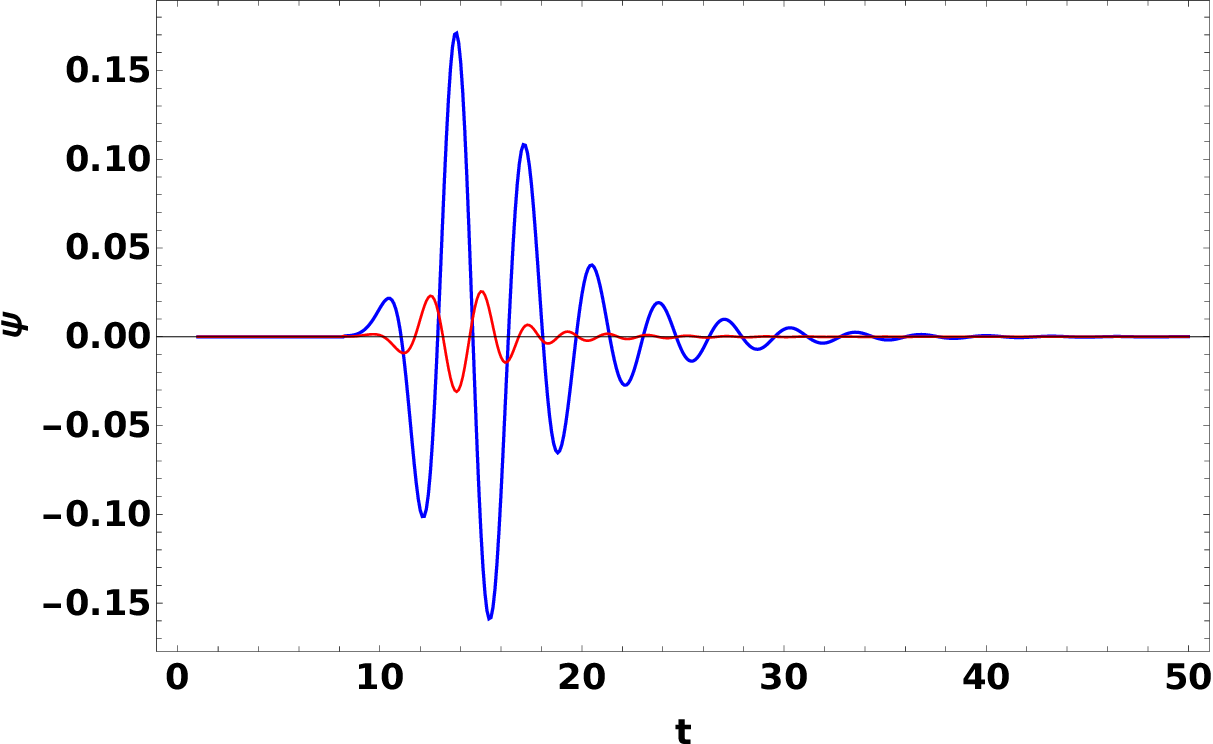}
    \caption{Comparison of echo obtained from subtracting single barrier effect for $n=10$ wormhole but with different $m$; $m=2$(blue) and $m=3$(red).}
    \label{fig:m3}
\end{figure}

\noindent $\bullet$ \textbf{How does echo signal change with increasing $m$?}\\
As $m$ increases for a particular $n$, the symmetric peaks on both sides of $\ell=0$ will be reducing in height and finally merging with the central peak to give a single barrier for high $m$. As a result the echo amplitude also decreases. 
Fig.(\ref{fig:m3}) shows how small the echo signal is for $m=3$ case in comparison to the $m=2$ when we just subtract the single barrier. These behaviours are as expected because of the small height of potential peaks.

\noindent From all the above observations regarding the echo structures we can come to the conclusion that the time evolution of a signal in the background of our wormhole geometry is much more diverse and complex because of the triple potential barrier. 

\section{Possible black hole mimicker?}

\begin{figure}[h]
 \centering
\begin{subfigure}[t]{0.4\textwidth}
  \centering
	\includegraphics[width=1\textwidth]{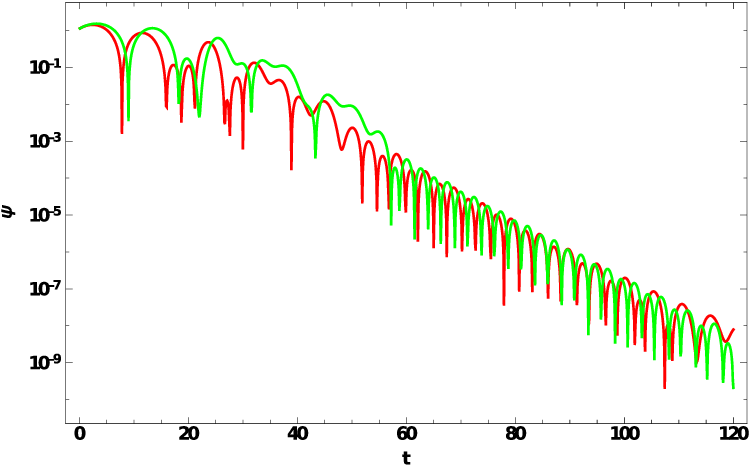}
 	\caption{ $n=2,m=3$}
 \end{subfigure}
 \hspace{1.5in}
\begin{subfigure}[t]{0.38\textwidth}
  \centering
	\includegraphics[width=1\textwidth]{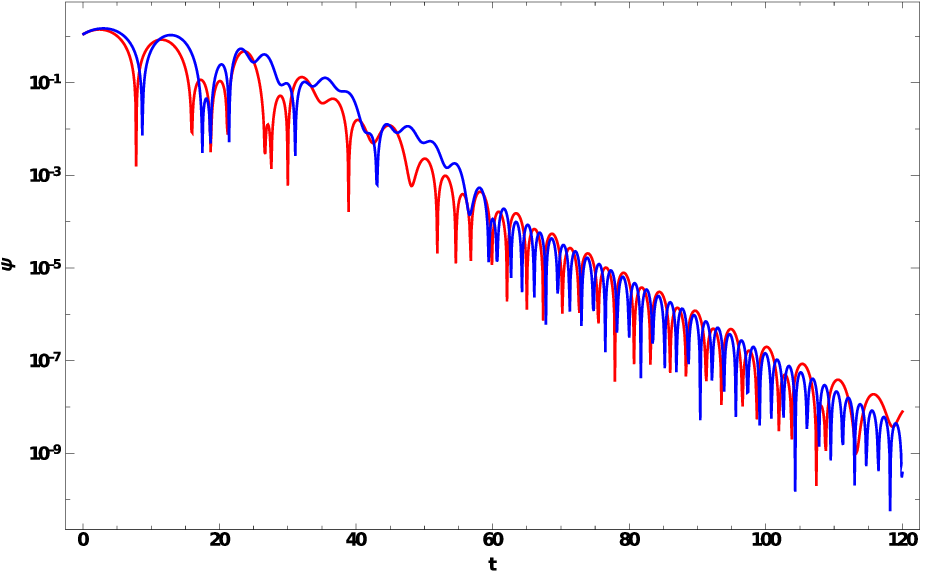}
 	\caption{ $n=4,m=3$}
 \end{subfigure}
 \hspace{1in}
 \begin{subfigure}[t]{0.38\textwidth}
 	\centering
 	\includegraphics[width=1\textwidth]{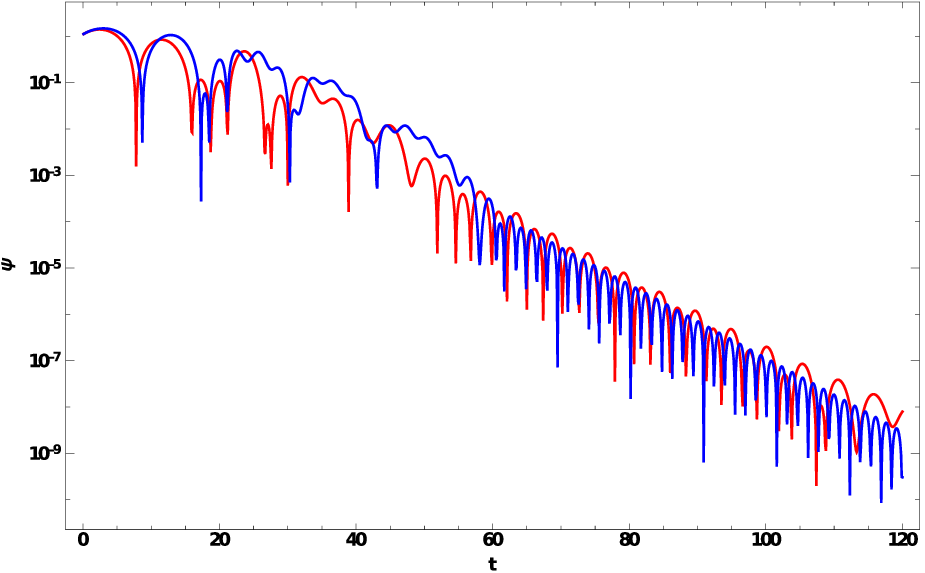}
 	\caption{$n=6,m=3$}
 	\end{subfigure}
 	\caption{Mimicking Schwarzschild BH ringing for m=3 angular mode; red plot denotes Schwarzschild ringing and blue and green denotes that of wormhole geometry.}
 	\label{fig:mimic}
 \end{figure}

\noindent One of the main motivations behind the study of our wormhole geometry is to develop 
the wormhole family as a potential black hole mimicker model. 
After the detection of gravitational waves, interest in studying  observable features of black holes have increased tremendously
-- in particular, with reference to
post-merger ringdown behavior via quasinormal modes.
Parallely, one notices an increase in studies on black hole mimickers which include a variety of compact objects sans the event horizon. These objects are expected to mimic some features of a traditional black hole and thus question the very existence and detection of the black hole through observations. It is therefore of utmost importance to study all possible mimicker models, not only to verify the existence of black holes without doubt but in this process also investigate any new physics that may arise. Having said that, any spacetime to exist in nature and mimic a black hole signature must pass through all possible stability analysis tests. The instability of some wormhole geometries under radial perturbation is well studied in literature \cite{shinkai_2002,gonzalez_2008,gonzalez_2008_1,bronnikov_2012}. So apart from the scalar wave propagation \cite{pdr_2020} and axial perturbation already performed that confirm the stability of our wormhole family, other perturbation tests need to be explored in future studies to ensure the viability of the mimicker model.\\
Wormholes are one of the contenders which have been studied as a mimicker in many works \cite{konoplya_2016,cardoso_2016,damour_2007,cardoso_2019}. Through our wormhole family metric we can simultaneously study the possibility of all the member wormholes being mimickers. We focus on the quasinormal ringing of the wormholes and try to find parameter values for which the ringing of the wormholes will be closest to that of a black hole in a similar vein, as done in \cite{konoplya_2016}. For the stability analysis, we had kept the throat radius parameter $b_0$ as unity. Now we will tweak it as well as `n'. We take the case of a Schwarzschild black hole with $2M=1$ so that for $m=3$ the fundamental QNM frequency is $1.19882 - i 0.185822$. We aim to find wormhole geometries in our family that might have similar QNM frequencies. Following are some scenarios where we find the closest mimicking behavior of the wormholes where the listed fundamental QNM values have been calculated using the Prony extraction scheme,
\begin{align*}
    n=2,\,m=3,\,b_0=2.16 \hspace{0.3in} & QNM: 1.36685 -i 0.189807\\
    n=4,\,m=3,\, b_0= 1.75  \hspace{0.3in} & QNM: 1.79846 - i 0.18086\\
    n=6,\,m=3,\,b_0=1.55 \hspace{0.3in} & QNM: 2.05558 -i 0.181839
\end{align*}
For the above mentioned wormhole geometries with specific throat radius, the dominant damping rate is very similar to that of the black hole but the corresponding oscillation frequency i.e. the real part of the QNM frequency is quite different. So the frequency of the QNM ringing can be used to distinguish any black hole mimicker from an actual black hole provided we observe the $m=3$ mode. As value of $n$ increases, it becomes easier to distinguish the wormholes from a black hole as the real part of the QNM increases. Fig.(\ref{fig:mimic}) shows the time domain signal for the above parameter values of the wormhole geometries in comparison to the behavior of the black hole. The damping rate is similar but the ringing structure determined by the real part of QNM frequency is quite different.\\
 
\begin{figure}
    \centering
    \includegraphics[width=0.4\textwidth]{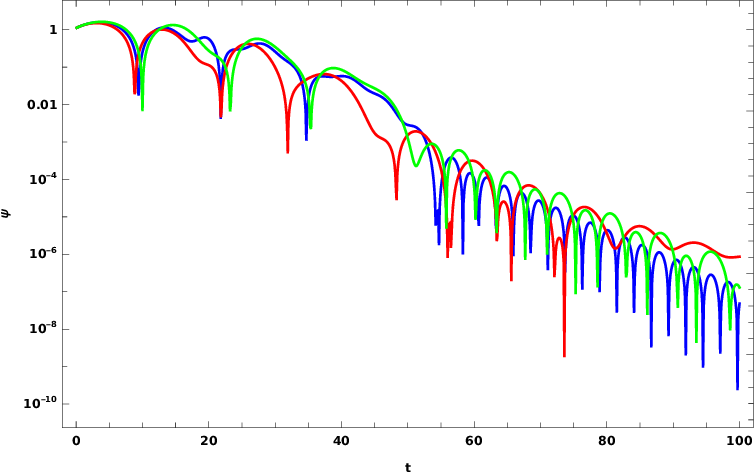}
    \caption{Schwarzschild profile (Red), $n=2$ (Green), $n=4$ (Blue) with the $b_0$ values mentioned above.}
    \label{fig:mimic1}
\end{figure}

\noindent The above mimicking properties of the wormholes hold true for the $m=3$ angular momentum mode. For mode $m=2$, the period over which  QNM ringing occurs in the time domain evolution plots for Schwarzschild BH is much less than that of wormholes even though the dominant QNM frequencies can be obtained to be similar for proper choice of throat radius (see fig.(\ref{fig:mimic1})). Thus, while for Schwarzschild black hole with $2M=1$ we have  $0.747 -i 0.1779$ as the dominant QNM frequency; for wormholes it becomes $0.976232 -i 0.171425$ corresponding to $n=2,m=2,b_0=1.78$ and for $n=4,m=2,b_0=1.54$, it is $1.20934 -i 0.174385$.

\noindent From the above analysis 
we come to the conclusion that our wormhole geometries can mimic only the damping rate of the ringdown profile of a black hole under certain parameter choices. The real part of the QNM will be quite distinct for the wormhole giving away their identity. On the other hand, higher $n$ wormhole geometries are poor mimickers and can easily be distinguished from their axial quasi-normal mode frequencies.

\section{\bf Discussion}

\noindent Our work comprises of a complete analysis of the behaviour of a two parameter Lorentzian wormhole family under axial gravitational perturbation. Following Chandrasekhar we calculated the master radial equation where an axially symmetric non-stationary spacetime represents the perturbed metric. The first significant result of this study is the observation of a triple potential barrier corresponding to lower angular momentum modes in $n>4$ geometries. As the mode number increases, the peaks come closer and finally merge into a single barrier. For $n=2$ and 4 wormholes, the potential is always a single barrier for all angular momentum modes. This is followed by the calculation of QNM frequencies for different wormhole geometries. The signal frequency was increasing as we went to higher angular momentum modes for all values of $n$. However, the behavior of the imaginary part of the QNMs that controls the damping rate is rather interesting. We observe long-lived modes which correspond to low damping rates as we go to higher angular momentum values in the $n>4$ wormholes. For $n=2$ case the imaginary part of QNM increases with increasing $m$ value making $m=2$ the lowest damped mode. Such long-lived modes might lead to instability of the spacetime under non-linear perturbation \cite{cardoso_2014} which needs to be explored in future. These modes can arise in different wormhole spacetimes as well which have been studied extensively in the literature \cite{churilova_2020,konoplya_2010}.
One of our aim of distinguishing the `shape' (geometry) of the wormholes just from their fundamental QNM frequency is also achieved as each geometry has a distinct frequency evident from the plots shown in fig.(\ref{fig:QNM_plot}) corresponding to a particular throat radius.\\
The multi-peak nature of the potential for our wormhole family makes it appropriate for studying the generation of gravitational wave echoes. The formation of echoes is much more involved and rich for a triple barrier potential as compared to a double barrier. The absence of sharpness in the potential peaks and the peaks being very closely spaced makes the echoes feeble and hard to see directly in the time domain profile. As a remedy, we try to `clean' the spectrum of the effects of scattering of signal from each barrier peak. First, we remove the effect due to the single barrier, then move on to the double barrier and finally remove the effect of scattering from the first well. This subtraction procedure leaves us with the net scattering off the entire triple barrier a.k.a the echoes. The echo profiles are observed and compared after all the subtraction schemes mentioned above. Even after such a cleaning procedure we are able to observe only a single echo signal because the amplitude of the signal gets damped very quickly making observation of multiple echoes rather impossible. We compare the strength of the echo profile for different parameter values as well as angular momentum modes. As expected, the strength of the echo signal decreases with an increasing value of $m$ because the potential peaks are slowly merging to form single barrier. Another interesting aspect regarding the stability of the spacetime is the presence of well in the potential whose depth increases as we go to higher $n$ geometry wormholes. The presence of well which may support bound states indicates possible instability. But for all wormhole geometries belonging to our family, we get a damped signal in the time domain evolution hence suggesting stability under linear axial perturbation.\\
We also briefly study the possibility of our wormhole family being a black hole mimicker by comparing the ringdown signal with that of a black hole. However, before exploring the mimicking features of our wormholes, it must be borne in mind that many wormhole geometries studied in literature do show instability under radial perturbation. Our wormhole family has narrowly escaped the instability scenario under axial perturbation despite the presence of `potential well' and `long-lived' modes. Thus, to ensure the robustness of our family, we need a complete stability analysis including radial perturbations, which we intend to pursue and complete in future. Keeping aside the above comments for now and focusing on our present work, we note that some parameter values can yield QNM frequencies that are very close to that of a black hole.  However even though the damping rate is nearly identical, the real part of the QNM is distinct and distinguishable from that of the black hole. So even though we get identical damping rates, the frequency would give away the identity of the wormholes when compared to a black hole. The larger $n$ wormhole geometries are even poor mimickers and can easily be distinguished using the axial QNMs. Future works can focus on the polar perturbations of the wormhole family and an even more accurate study would be to include rotation in the metric, since, in nature, all astrophysical objects are known to be rotating.
In summary, we performed a detailed analysis of the QNM spectrum and the echo structure for our wormhole family. We have also speculated on how such wormholes can be of use as black hole mimicker candidates. \\

\noindent {\bf Acknowledgments}

\noindent The author is grateful to Prof. Sayan Kar for suggesting the problem, carefully reading the manuscript and for his valuable comments in improving it. She also thanks Indian Institute
of Technology, Kharagpur, India for support and for allowing  her to use all available facilities there.

\appendix

\section{Perturbation of energy-momentum tensor }

\noindent Axial perturbation of the metric is associated with inducing rotation of the perturbed object due to presence of non-zero cross-terms in the metric as shown in eq.(\ref{eq:pert_metric}). Such a perturbation will affect both the metric as well as the matter content of the spacetime. While deriving the perturbation equation we follow the notation used in \cite{chandrasekhar_1985} with the coordinates being $t (=x^0)$, $\phi(=x^1)$, $r(= x^2)$ and $\theta(=x^3)$. The Einstein's equation for the perturbed metric of eq.(\ref{eq:pert_metric}) will be 
\begin{equation}
     \bar{G}_{\mu \nu} = T_{\mu \nu} + \delta T_{\mu \nu}
\end{equation}
where the total energy-momentum tensor is taken as a sum of the background contribution $T_{\mu \nu}$ and a small perturbation to it. The metric of eq.(\ref{eq:pert_metric}) can be written in tetrad basis as $ds^2 = \eta_{a b} e^a e^b$ with  $\eta_{ab}$ being the Minkowskian metric and $a, b$ denoting the tetrad indices. Hence the corresponding tetrads will be of the form
\begin{align}
     & e^0_{\mu} = ( e^\nu , 0, 0, 0) \\
   & e^1_\mu =  ( -\sigma e^\psi, e^{\psi}, - q_r e^\psi,  -q_\theta e^\psi)\\
    & e^2_\mu = (0, 0, e^{\mu_r}, 0) \\
    & e^3_\mu = (0, 0, 0, e^{\mu_\theta}) 
\end{align}
By setting $q_r = q_\theta =\sigma =0$ the tetrads for the background unperturbed metric will be obtained.
For axial perturbation, we will be interested in $\phi t$, $\phi r$ and $\phi \theta$ components of the perturbation equation which will correspond to $1 0$,$1 2$ and $1 3$ components in frame basis. Thus we get
\begin{equation}
    \bar{R}_{ab} = \delta T_{ab}
\end{equation}
as $\eta_{10}= \eta_{12} = \eta_{13} = 0$ and $T_{10} = T_{12} = T_{13}=0$. To complete the analysis, the perturbation in the energy-momentum tensor needs to be calculated. In case of a known matter field sourcing the geometry, the energy-momentum tensor can be directly perturbed to find the RHS of the perturbation equation. It can be seen that for a spacetime sourced by a scalar field,  $\delta T_{ab} =0$ however for solutions of Einstein-Maxwell systems $\delta T_{ab} \neq 0$ like in the Reissner-Nordstrom black hole \cite{chandrasekhar_1985} (see also \cite{bronnikov_2012} for example in a wormhole spacetime). To continue a similar analysis in absence of an underlying matter theory sourcing our $n>2$ wormhole geometries, we follow \cite{chen_2019} and consider the effective energy-momentum tensor as obtained from Einstein's equation 
\begin{align}
& \rho (\ell) = -2 \frac{r''}{r} - \left (\frac{r'}{r}\right )^2+\frac{1}{r^2}\\
& \tau (\ell) = \left (\frac{r'}{r}\right )^2 - \frac{1}{r^2} \\
& p(\ell) = \frac{r''}{r}
\label{eq:EM_tensor}
\end{align}
with $r(\ell) = (\ell^n + b_0^n)^{1/n}$. Note that the Ellis-Bronnikov wormhole corresponding to $n=2$ case is supported by a phantom massless scalar field of the form $\phi(\ell/b_0) = \sqrt{2}\, tan^{-1}(\ell/b_0)$. Again, for a phantom scalar field, $\rho_\phi = \tau_\phi = -p_\phi = -(1/2)\phi'^2$ where the derivative is with respect to $\ell$. Thus the above equations with $r(\ell) = \sqrt{\ell^2+b_0^2}$ are indeed satisfied by the scalar field giving,
\begin{align}
& \rho (\ell) = -2 \frac{r''}{r} - \left (\frac{r'}{r}\right )^2+\frac{1}{r^2} = \frac{-b_0^2}{(\ell^2+ b_0^2)^2} = \frac{-1}{2} \phi'^2\\
& \tau (\ell) = \left (\frac{r'}{r}\right )^2 - \frac{1}{r^2}= \frac{-b_0^2}{(\ell^2+ b_0^2)^2} = \frac{-1}{2} \phi'^2 \\
& p(\ell) = \frac{r''}{r}= \frac{b_0^2}{(\ell^2+ b_0^2)^2} = \frac{1}{2} \phi'^2
\label{eq:EM_tensor1}
\end{align}
These have been discussed in detail in Paper-I along with properties of energy-momentum tensor satisfied by $n>2$ wormholes. \\
Now, the energy-momentum tensor, as shown in eq. A7, A8 and A9, is interpreted as an anisotropic fluid of the form
\begin{equation}
 T_{\mu \nu} = (\rho + p) u_\mu u_\nu + (\tau -p) x_\mu x_\nu + p g_{\mu \nu}
\end{equation}
where $g_{\mu \nu}$ corresponds to the background metric and $\rho, \tau, p$ are the energy density, radial and tangential pressure respectively. $u^\mu (= u^t,0,0,0)$ and $x^\mu(=0,x^r,0,0)$ are  respectively timelike four-velocity and spacelike unit vector orthogonal to $u^\mu$. Converting to frame basis and taking small perturbation we arrive at
\begin{equation}
    \delta T_{ab} = \delta(\rho + p) u_a u_b + (\rho + p) \delta(u_a u_b) + \delta(\tau -p) x_a x_b + (\tau -p) \delta(x_a x_b) + \delta p \,\eta_{a b}.
\end{equation}
It can now be easily visualized that $\delta T_{10}= \delta T_{12} =\delta T_{13} =0$ which leads to the final perturbation equation being the one shown in eq.(\ref{eq:perturbation}) with the RHS being 0.

\section{Consistency of perturbation equations}

\noindent As mentioned earlier, axial perturbation is associated with $(t \phi), (r \phi)$ and $(\theta \phi)$ components of the perturbation equation. We have used only the $(r \phi)$ and $(\theta \phi)$ components to arrive at the master equation \ref{eq:radial_eq}. So it is important to ensure that the $\bar{R}_{01}$ or $\bar{R}_{t \phi }$ component is consistent with the other two equations and does not give any new constraints on the system. Following Chandrasekhar \cite{chandrasekhar_1985}, from $\bar{R}_{r \phi}=0$ we get,
\begin{equation}
    i\, \omega\, Q_{t r} = \frac{1}{r^2} (3\, cot\theta\,\, Q_{r \theta} + Q_{r \theta,\theta} )
    \label{eq:q1}
\end{equation}
where $Q_{t r} = \sigma_{,r} -q_{r,t}$ and $\omega$ appears after incorporating time dependence. Similarly, $\bar{R}_{\theta \phi}=0$ gives 
\begin{equation}
    \frac{-r^2 i \omega Q_{t \theta}}{\sqrt{1-\frac{b(r)}{r}}} = \Big[2 r \sqrt{1- \frac{b(r)}{r}} + \frac{r^2}{2 \sqrt{1- \frac{b(r)}{r}}} \Big(\frac{- b'(r)}{r} +\frac{b(r)}{r^2} \Big)\Big] Q_{r \theta} + r^2 \sqrt{1-\frac{b(r)}{r}} Q_{r \theta, r}
    \label{q2}
\end{equation}
with $Q_{r \theta} = q_{r,\theta} - q_{\theta,r}$. Now, the form of $\bar{R}_{t \phi}$ without the overall common factors become
\begin{equation}\begin{split}
  \bar{R}_{t \phi}= \Big[ 4 r^3 \sqrt{1- \frac{b(r)}{r}} + \frac{r^4}{2 \sqrt{1- \frac{b(r)}{r}}} \Big(\frac{-b'(r)}{r} + \frac{b(r)}{r^2} \Big) \Big] Q_{t r} + r^4 \sqrt{1- \frac{b(r)}{r}} Q_{tr,r} +\\
   3 cot\theta \frac{r^2}{\sqrt{1-\frac{b(r)}{r}}} Q_{t \theta} + \frac{r^2}{\sqrt{1-\frac{b(r)}{r}}} Q_{t \theta,\theta}
   \end{split}
\end{equation}
Substituting $Q_{t r}\,,\, Q_{t \theta}$ from eq.(\ref{eq:q1}) and eq.(\ref{q2}) and $\, Q_{tr,r}, \, Q_{t \theta,\theta}$ from the derivatives of the equations, we find $\bar{R}_{t \phi}$ to be trivially 0. Hence, the three axial perturbation equations are consistent for any general form of $b(r)$.
\bibliography{ref}
\end{document}